\documentclass[aps,prl,twocolumn,groupedaddress,showpacs]{revtex4}

\usepackage{bm}
\usepackage{graphicx}

\begin{document}

\title{Superconducting gap symmetry of Ba$_{0.6}$K$_{0.4}$Fe$_2$As$_2$ studied by angle-resolved photoemission spectroscopy}

\author{
	K. Nakayama,$^1$
	T. Sato,$^{1,2}$
	P. Richard,$^3$
	Y.-M. Xu,$^4$
	Y. Sekiba,$^1$
	S. Souma,$^3$
	G. F. Chen,$5$
	J. L. Luo,$^5$
	N. L. Wang,$^5$
	H. Ding,$^5$
	and T. Takahashi$^{1,3}$
	}
	
\affiliation{$^1$Department of Physics, Tohoku University, Sendai 980-8578, Japan}
\affiliation{$^2$TRIP, Japan Science and Technology Agency (JST), Kawaguchi 332-0012, Japan}
\affiliation{$^3$WPI Research Center, Advanced Institute for Material Research, Tohoku University, Sendai 980-8577, Japan}
\affiliation{$^4$Department of Physics, Boston College, Chestnut Hill, MA 02467, USA}
\affiliation{$^5$Beijing National Laboratory for Condensed Matter Physics, and Institute of Physics, Chinese Academy of Sciences, Beijing 100080, China}

\date{\today}

\begin{abstract}
We have performed high-resolution angle-resolved photoemission spectroscopy on the optimally-doped Ba$_{0.6}$K$_{0.4}$Fe$_2$As$_2$ compound and determined the accurate momentum dependence of the superconducting (SC) gap in four Fermi-surface sheets including a newly discovered outer electron pocket at the M point.  The SC gap on this pocket is nearly isotropic and its magnitude is comparable ($\Delta$ $\sim$ 11 meV) to that of the inner electron and hole pockets ($\sim$12 meV), although it is substantially larger than that of the outer hole pocket ($\sim$6 meV).  The Fermi-surface dependence of the SC gap value is basically consistent with $\Delta$($k$) = $\Delta$$_0$cos$k_x$cos$k_y$ formula expected for the extended $s$-wave symmetry.  The observed finite deviation from the simple formula suggests the importance of multi-orbital effects.
\end{abstract}

\maketitle

Understanding the nature of the superconducting (SC) gap in a superconductor is critically important in establishing the microscopic origin of superconductivity since the magnitude and the symmetry of the SC gap are directly associated with the pairing strength and the pairing interactions, respectively.  The recent discovery of iron-based superconductors with $T_{\rm c}$ values up to 55 K \cite{Kamihara, SmNature, AIST, Ba122} has generated fierce debates on its high-$T_{\rm c}$ mechanism.  In the iron-based superconductors, all the Fe 3$d$-derived bands are located in the vicinity of the Fermi level ($E_{\rm F}$), leading to the emergence of multiple Fermi surface (FS) sheets as predicted by band structure calculations \cite{Lebegue, Singh, Xu}.  It is thus particularly important to elucidate the accurate FS sheet dependence of the SC gap to gain insight into the role of different orbitals (FSs) to the pairing mechanism.  Angle-resolved photoemission spectroscopy (ARPES) is a powerful technique to directly observe the momentum-resolved SC gap.  Recent high-resolution ARPES studies on hole-doped Ba$_{0.6}$K$_{0.4}$Fe$_2$As$_2$ have revealed FS-dependent nodeless superconducting gaps \cite{HongEPL, Zhou}.  The observed same and anomalously large SC gap on two different small FS sheets, which are connected by the antiferromagnetic wave vector \cite{HongEPL}, suggested the importance of inter-band interactions for the pairing \cite{Mazin, Kuroki, Lee}.  Although these results provide important insights in understanding the SC mechanism, a full story regarding the pairing mechanism is still unresolved, and in fact, there are several remaining important issues yet to be clarified: (i) LDA (Local Density Approximation) calculations \cite{Xu} predict the presence of two electron pockets at the M point in the Brillouin zone, whereas the previous ARPES studies revealed only one \cite{HongEPL, Kaminski122}.  (ii) It is unclear whether or not the FS sheets connected $via$ the antiferromagnetic wave vector displays exactly the same SC gap magnitude.  (iii) The presence or the absence of a pseudogap above $T_{\rm c}$ is highly controversial among ARPES experiments \cite{HongEPL, Zhou, Kaminski}.  (iv) The microscopic origin of the orbital dependence of the SC gap is still beyond complete understanding.

In this Letter, we report the observation of a second electron pocket centered around the M point.  We give the detailed momentum dependence of the SC gap in Ba$_{0.6}$K$_{0.4}$Fe$_2$As$_2$ ($T_{\rm c}$ = 37 K) on each of the four Fermi surface sheets.  The results indicate distinct difference in the SC gap size among different FSs, while the SC gap within the same FS sheet is fairly isotropic.  We discuss the implications of these results in relation to inter-band scattering and the model gap function with the extended $s$-wave symmetry.

\begin{figure}[!t]
\includegraphics[width=8cm]{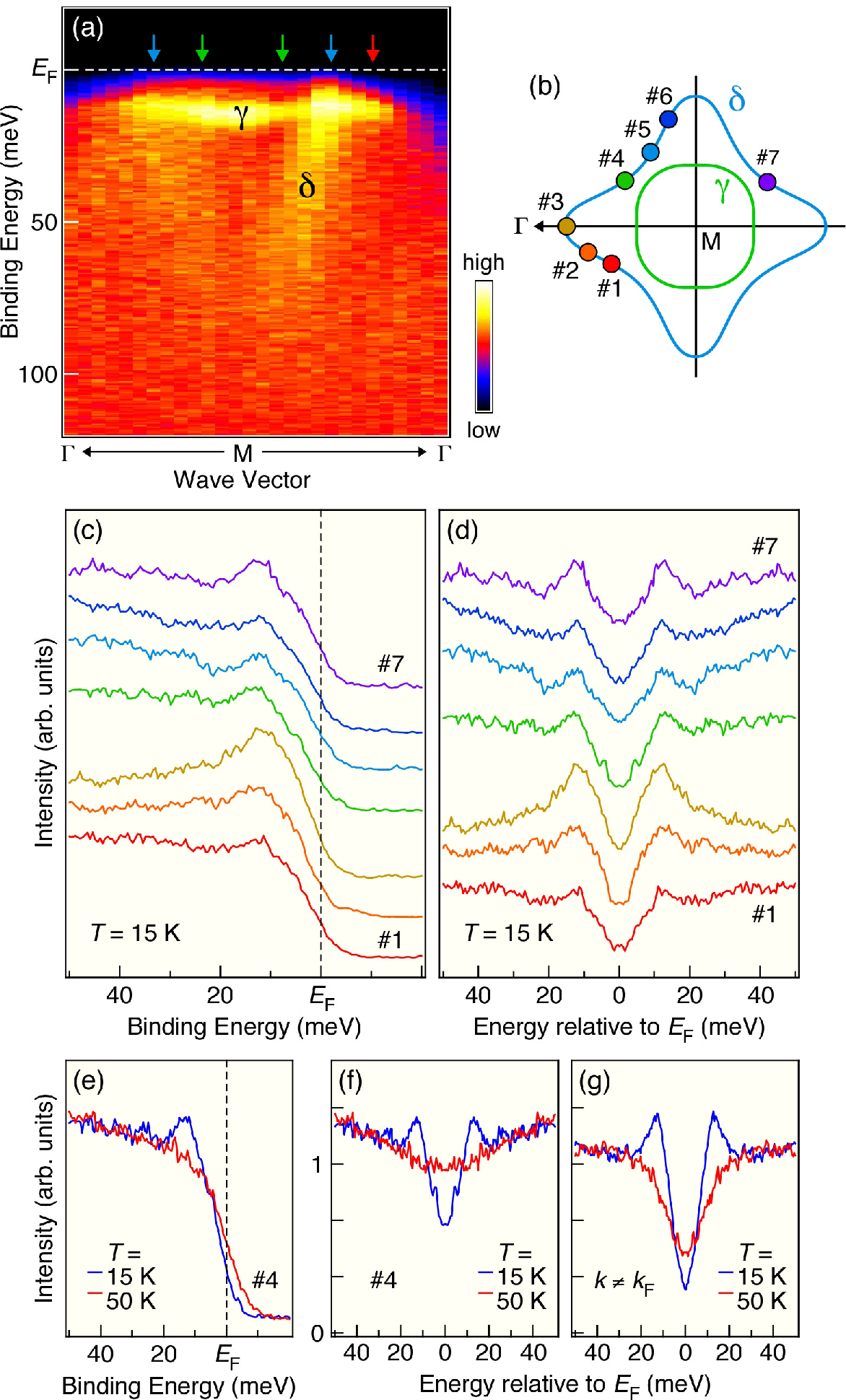}
\caption{(a) ARPES intensity of Ba$_{0.6}$K$_{0.4}$Fe$_2$As$_2$ ($T_{\rm c}$ = 37 K) as a function of binding energy and wave vector measured at 15 K along the $\Gamma$-M high symmetry line around the M point.  Green and light blue arrows show the $k_{\rm F}$ positions of the $\gamma$ and $\delta$ electron pockets, respectively.  (b) Schematic FSs of the $\gamma$ (green curve) and $\delta$ (light blue curve) pockets.  The wave vector of the M point corresponds to ($\pi$, 0) in the unfolded Brillouin zone.  (c) Representative ARPES spectra at 15 K measured at several $k_{\rm F}$ points ($\sharp$1-7) of the $\delta$ FS shown by circles in (b).  The coloring of the spectra is the same as that of the circles in (b).  (d) Symmetrized spectra of (c).  (e) and (f), Temperature dependence of (e) the ARPES spectrum and (f) its symmetrized counterpart at $k_{\rm F}$ point $\sharp$4.  (g) Temperature dependence of an ARPES spectrum measured slightly away from $k_{\rm F}$ point as denoted by red arrow in (a).}
\label{Fig.1}
\end{figure}

The high-quality single crystals of Ba$_{0.6}$K$_{0.4}$Fe$_2$As$_2$ ($T_{\rm c}$ = 37 K) used in this study were grown by the flux method \cite{Chen}.  High-resolution ARPES measurements were performed using a VG-SCIENTA SES2002 spectrometer with a high-flux He discharge lamp and a toroidal grating monochromator.  We used the He I$\alpha$ resonance line ($h\nu$ = 21.218 eV) to excite photoelectrons.  The energy and angular resolutions were set at 2-4 meV and 0.2$^{\circ}$, respectively.  Sample orientations were determined by Laue x-ray-diffraction patterns prior to the ARPES measurements.  Clean surfaces for ARPES measurements were obtained by cleaving the crystals $in$ $situ$ in a working vacuum better than 5$\times$10$^{-11}$ Torr.  The Fermi level ($E_{\rm F}$) of the samples was referenced to that of a gold film evaporated onto the sample substrate.  Low-energy electron diffraction on the measured surface shows a sharp 1$\times$1 pattern without any detectable reconstruction down to 80 K.  A mirror-like sample surface was found to be stable without obvious degradation for typical measurement periods of 3 days.

Figure 1(a) shows the ARPES spectral intensity of Ba$_{0.6}$K$_{0.4}$Fe$_2$As$_2$ measured at 15 K along the $\Gamma$-M high-symmetry line.  We clearly find two electron-like band dispersions centered at the M point \cite{HongBandstructure}, in agreement with LDA band calculations \cite{Xu}.  This contrasts with previous ARPES studies that could clearly distinguish only one electron-like pocket centered at the M point \cite{HongEPL, Kaminski122}.  The FSs of the inner shallow pocket (called $\gamma$ band) and the outer deeper pocket ($\delta$ band) have a circular-like and a cross-like shape, respectively (Fig. 1(b)).  The existence of these two FSs is basically explained by the hybridization of two ellipsoidal pockets elongated along two $\Gamma$-M directions (along $k_x$ and $k_y$ axes) \cite{HongBandstructure}.  In the SC state, both bands show a local minimum at their corresponding Fermi vectors ($k_{\rm F}$) (green and light blue arrows in Fig. 1(a)) and disperse back toward higher binding energy by following a Bogoliubov-quasiparticle-like dispersion, indicating the opening of the SC gap.  Here we focus on the SC gap of the newly discovered $\delta$ band.  Figure 1(c) shows the near-$E_{\rm F}$ ARPES spectra at various $k_{\rm F}$ points of the $\delta$ band at 15 K.  We clearly observe a sharp quasiparticle peak together with a leading-edge shift in all the spectra, suggesting that the SC gap opens in the entire momentum region on the $\delta$ FS.  A weak shoulder-like feature is observed at around 6 meV, as also observed in the $\alpha$ and $\gamma$ FS pockets \cite{HongEPL}.  It is noted that this shoulder, although its origin is unknown, does not affect the determination of the SC gap size, since the SC quasiparticle peak is much pronounced and its energy position is well identified.  In order to estimate the SC gap size at each $k_{\rm F}$ point, we have symmetrized the ARPES spectrum to eliminate the effect of the Fermi-Dirac distribution function \cite{Norman}.  As clearly visible in Fig. 1(d), the separation of two quasiparticle peaks, which corresponds to the full SC gap size (2$\Delta$), does not show significant momentum dependence, suggesting a nearly isotropic SC gap opening.  Interestingly, the estimated gap size ($\Delta$ $\sim$ 11 meV) is approximately twice as large as the expected value from the weak coupling BCS theory (5.6 meV; 2$\Delta$/$k_{\rm B}$$T_{\rm c}$ = 3.52), demonstrating the strong-coupling character of the superconductivity on the $\delta$ band.

To address the presence or absence of pseudogap \cite{HongEPL, Zhou, Kaminski}, we have measured ARPES spectra above $T_{\rm c}$.  Figures 1(e) and 1(f) show the temperature dependence of the $k_{\rm F}$ spectrum ($\sharp$4) and its symmetrized counterpart, respectively.  It is apparent from the 50-K spectrum that the sharp quasiparticle peak disappears and the midpoint of the leading edge is located at $E_{\rm F}$, showing no clear evidence for pseudogap opening, unlike a previous ARPES study on the same compound reporting a strong pseudogap effect \cite{Zhou}.  We caution that one must be careful about the determination of the $k_{\rm F}$ position, since even a small miss-assignment of the $k_{\rm F}$ position may cause a false pseudogap-like behavior in the ARPES spectrum as shown in Fig. 1(g).  The presence of the flat $\beta$ band just below $E_{\rm F}$ around the M point \cite{HongBandstructure} can easily induce such miss-assignment.

\begin{table}
\begin{center}
{\footnotesize
{\tabcolsep=0.1cm
\begin{tabular}{|c||c|c|c|c|c|c|c|}
\hline
\multicolumn{1}{|c||}{} & \multicolumn{1}{c|}{A} & \multicolumn{1}{c|}{B} & \multicolumn{1}{c|}{C} & \multicolumn{1}{c|}{D} & \multicolumn{1}{c|}{E} & \multicolumn{1}{c|}{F} & \multicolumn{1}{|c|}{G} \\ \hline\hline
$\alpha$ & 11.2-12.8 & 11.8-12.6 & 11.8-12.5 & 11.4-13.6 & 11.9 & 11.5-12.3 & - \\  \hline
$\beta$ & 4.9-5.4 & 4.6-5.4 & - & 5.5-7.5 & 5.1 & - & 5.0 \\  \hline
$\gamma$ & 11.8-12.4 & 12.7 & 11.6-12.7 & - & - & - & - \\  \hline
$\delta$ & 10.2-12.1 & 11.0-11.1 & 11.1-11.3 & - & - & - & - \\  \hline
\end{tabular}
}
}
\end{center}
\caption{FS dependence of the SC gap size (meV) in Ba$_{0.6}$K$_{0.4}$Fe$_2$As$_2$ measured at 15 K on seven different samples (A-G).  The $\alpha$ and $\beta$ ($\gamma$ and $\delta$) FSs correspond to the inner and outer hole-like (electron-like) FSs, respectively.  Distributions of the SC gap size arise from the differences in the estimated SC gap size among various $k_{\rm F}$ points on each FS sheet.}
\label{Table 1}
\end{table}

\begin{figure}[!t]
\includegraphics[width=8cm]{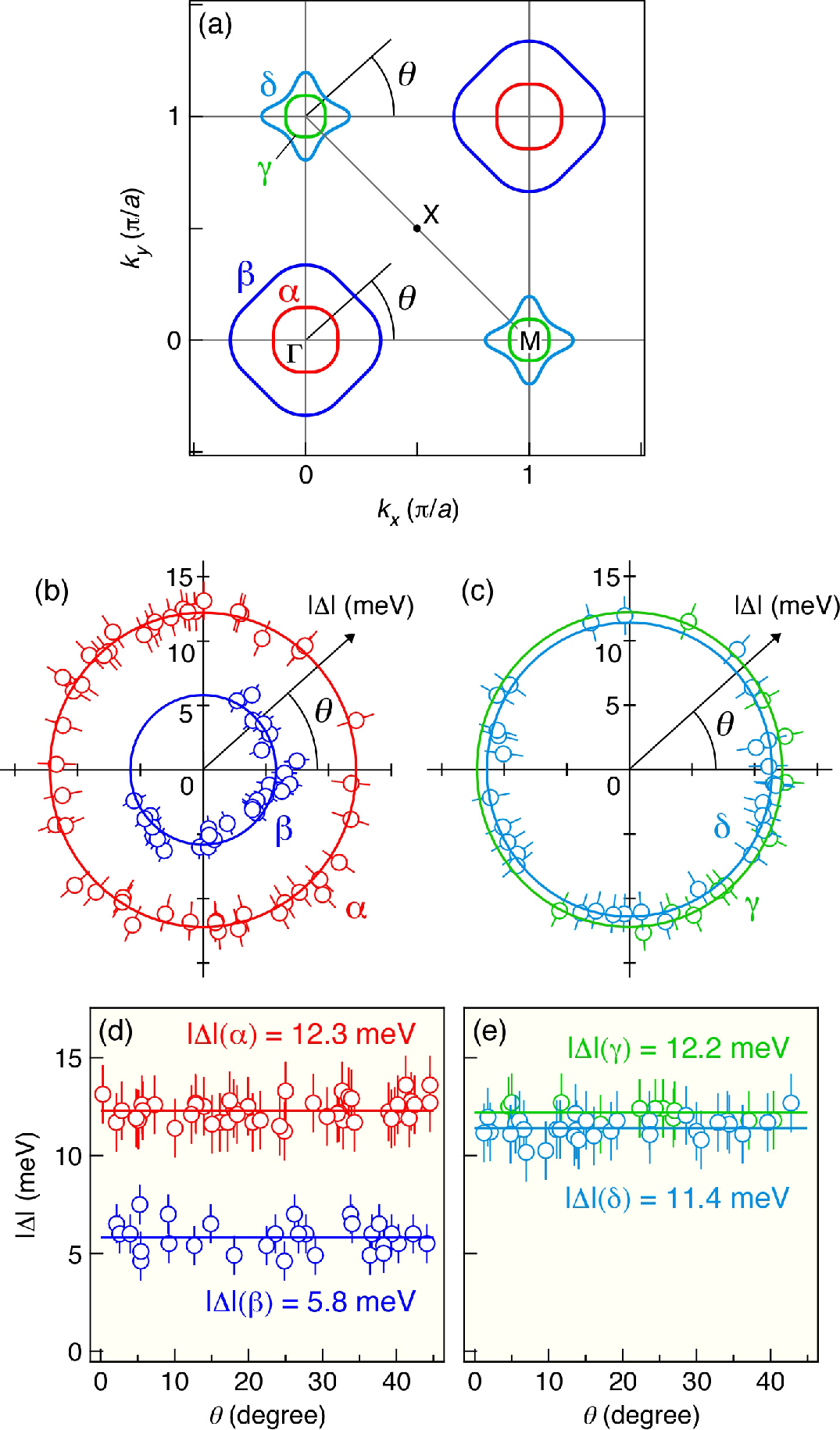}
\caption{(a) Schematic view of the four FS sheets with a definition of the FS angle ($\theta$).  (b), (c) SC gap size at 15 K extracted from the symmetrized spectra shown on polar plots for the (b) $\alpha$, $\beta$ and (c) $\gamma$, $\delta$ FSs as a function of $\theta$.  (d) and (e), Same as (b) and (c) but the data points have been shifted (reduced) into the 0$^{\circ}$ $\leq$ $\theta$ $\leq$ 45$^{\circ}$ region, by assuming a four-fold symmetry.  Thick lines show the averaged SC gap values on each FS.}
\label{Fig.2}
\end{figure}

Next we discuss the FS dependence of the SC gap.  Table 1 summarizes the ARPES results of our comprehensive SC gap measurements over the whole FS sheets for seven different samples (A-G).  The obtained gap sizes for each FS are almost constant among several different cleaved surfaces, demonstrating a high reproducibility.  The SC gap sizes on the $\alpha$, $\gamma$, and $\delta$ bands are comparable, while the SC gap on the $\beta$ band is significantly smaller.  The detailed momentum dependence of the SC gap on each FS is displayed on polar plots as a function of the FS angle ($\theta$) in Figs. 2(b) and 2(c).  As clearly visible, the SC gap is nearly isotropic and definitely nodeless on all the four FS sheets.  In Figs. 2(d) and 2(e), we plot all the data points shifted (reduced) into 0$^{\circ}$ $\leq$ $\theta$ $\leq$ 45$^{\circ}$, by assuming a four-fold symmetry.  Obviously, there is no systematic change of the gap size as a function of $\theta$ within the present experimental uncertainty, confirming the isotropic nature of the gap.  We obtained averaged gap sizes $\Delta$ of 12.3 $\pm$ 0.6, 5.8 $\pm$ 0.8, 12.2 $\pm$ 0.3, and 11.4 $\pm$ 0.5 meV for the $\alpha$, $\beta$, $\gamma$, and $\delta$ FS, respectively, corresponding to 2$\Delta$/$k_{\rm B}T_{\rm c}$ ratios of 7.7 $\pm$ 0.4, 3.6 $\pm$ 0.5, 7.7 $\pm$ 0.2, and 7.2 $\pm$ 0.3.  By referring to these values, the gap size of the $\delta$ band is slightly smaller than that of the $\alpha$ and $\gamma$ bands, implying the existence of three categories of SC gaps.

Now we discuss the implication of the observed FS-dependent SC gap in relation to the SC order parameter.  In our previous study \cite{HongEPL}, the large and almost same SC gap size on the well nested hole-like $\alpha$ and electron-like $\gamma$ FS sheets has been attributed to the enhanced pairing amplitude by inter-band scattering $via$ $Q$ $\sim$ ($\pi$, 0) spin fluctuations.  In this study, we additionally found that the newly discovered electron-like $\delta$ band shows a nodeless SC gap, whose magnitude is slightly smaller than that of the $\alpha$ and $\gamma$ bands.  The observed strong-coupling character of the $\delta$ band establishes the important role of inter-band interactions for the pairing \cite{HongEPL}, since the $\delta$ FS is also well connected to the $\alpha$ FS $via$ the $Q$ $\sim$ ($\pi$, 0) wave vector.  A pairing mediated by $Q$ $\sim$ ($\pi$, 0) spin fluctuations would lead to the opening of an unconventional sign-changing $s$-wave (extended $s$-wave) gap on the hole and electron pockets \cite{Mazin, Kuroki, Lee}, consistent with the observed nodeless SC gaps.  Our finding further suggests that the same gap amplitude ($\pm\Delta$) is not required between the well-connected FS sheets.  This may be related to different intra-band pairing strengths along different FSs.

\begin{figure}[!t]
\includegraphics[width=8cm]{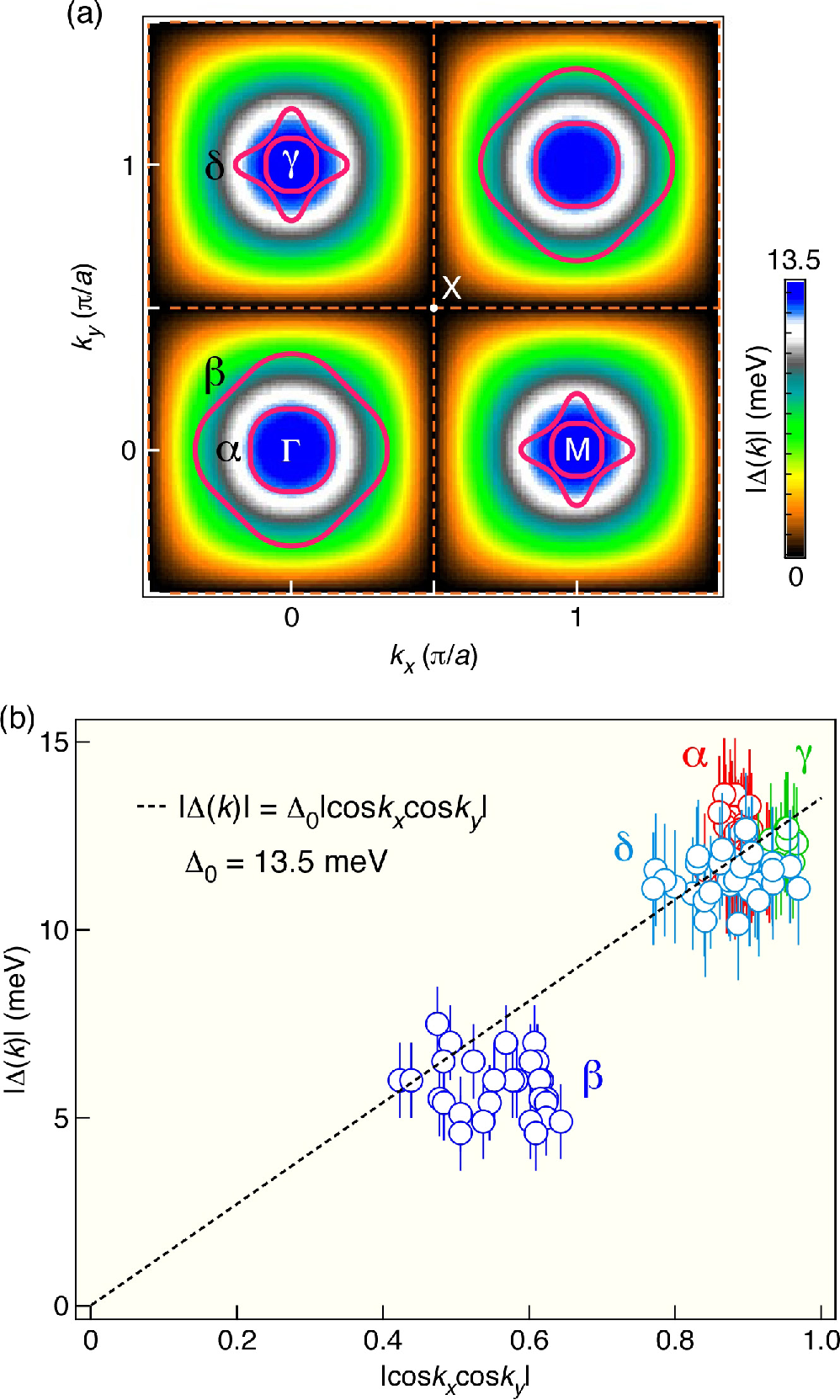}
\caption{(a) Theoretical SC gap value $\mid$$\Delta$($k$)$\mid$ = $\Delta$$_0$$\mid$cos$k_x$cos$k_y$$\mid$ with $\Delta$$_0$ = 13.5 meV as a function of the two-dimensional wave vector \cite{Seo, Korshunov}.  Dashed lines represent zero-gap (nodal) lines.  The schematic four FSs are superimposed (pink curves).  (b) Plot of the SC gap values at 15 K as a function of $\mid$coos$k_x$cos$k_y$$\mid$.  The dashed line shows the best fit using the gap function $\mid$$\Delta$($k$)$\mid$ = $\Delta$$_0$$\mid$cos$k_x$cos$k_y$$\mid$.}
\label{Fig.3}
\end{figure}

Since the full set of SC gaps as a function of momentum is now available, it is tempting to quantitatively compare the present results with existing theoretical models.  Here we examine the simple formula $\Delta$($k$) = $\Delta$$_0$cos$k_x$cos$k_y$, known as one of the SC gap form for the extended $s$-wave symmetry \cite{Seo, Korshunov}.  This formula predicts (i) the opening of larger (smaller) SC gap on smaller (larger) FS, (ii) a sign change between hole and electron FSs, and (iii) a full-gap opening on each FS.  We note that, although the cos$k_x$cos$k_y$ value becomes zero between the $\Gamma$ and M line (dashed lines in Fig. 3(a)), it does not create a gap node in the case of Ba$_{0.6}$K$_{0.4}$Fe$_2$As$_2$, since observed FSs do not intersect this nodal line.  In Fig. 3(b), we plot the experimentally determined gap values as a function of $\mid$cos$k_x$cos$k_y$$\mid$.   The observed gap sizes of the four FS sheets basically follow the $\Delta$$_0$$\mid$cos$k_x$cos$k_y$$\mid$ function with $\Delta$$_0$ = 13.5 meV.  However, a closer look reveals a finite deviation between the experiment and the model.  For example, according to the model, the gap size on the $\beta$ FS varies from 6 (along $\Gamma$-M) to 8.5 meV (along $\Gamma$-X), whereas the experimentally observed nearly isotropic gap value (5.8 $\pm$ 0.8 meV) does not follow this trend.  Similar deviation is also seen in the $\delta$ FS.  These results indicate that the observed FS-dependent SC gaps are not simply explained by a cos$k_x$cos$k_y$ order parameter with a single energy scale $\Delta$$_0$, suggesting that multi-orbital effects should be seriously taken into account to understand the pairing mechanism of the iron-based superconductors.  In addition, the inter-layer coupling may also change the gap function.

In summary, we have reported comprehensive ARPES results for the SC gap on four different FSs on optimally-doped Ba$_{0.6}$K$_{0.4}$Fe$_2$As$_2$ ($T_{\rm c}$ = 37 K), including the newly discovered electron-like $\delta$ FS.  We have found that the $\delta$ band shows the opening of a slightly smaller SC gap below $T_{\rm c}$ compared to another electron pocket (the $\gamma$ FS), but its magnitude is still in the strong-coupling regime, supporting the inter-band scattering scenario for the pairing mechanism.  Furthermore, our detailed measurements revealed that the momentum dependence of the SC gaps basically agree with the simple SC gap function $\Delta$$_0$cos$k_x$cos$k_y$, while there is a finite deviation suggesting the importance of multi-orbital effects.  The nearly isotropic but Fermi surface dependent character of the SC gap puts a strong constraint on theoretical models promoted to explain the superconducting mechanism in the iron-based superconductors.

We thank X. Dai, Z. Fang, J. P. Hu, and Z. Wang for thier valuable discussions and suggestions.  This work was supported by grants from JSPS, JST-TRIP, JST-CREST, MEXT of Japan, the Chinese Academy of Sciences, Ministry of Science and Technology of China, and NSF of US.

\end{document}